\documentclass[12pt]{iopart}

\usepackage{hyperref}
\usepackage{graphicx}
\usepackage{iopams}  
\usepackage[e]{esvect}

\begin{document}


\title[The UV-LED Mission]{Ground Testing and Flight Demonstration of Charge Management of Insulated Test Masses Using UV-LED Electron Photoemission}

\author{Shailendhar Saraf$^{1,4}$, Sasha Buchman$^1$, Karthik Balakrishnan$^1$, Chin Yang Lui$^1$, Michael Soulage$^2$, Dohy Faied$^2$, John Hanson$^2$, Kuok Ling$^2$, Belgacem Jaroux$^2$, Abdullah AlRashed$^3$, Badr Al Nassban$^3$, Badr Al Suwaidan$^3$, Mohammed Al Harbi$^3$, Badr Bin Salamah$^3$, Mohammed Bin Othman$^3$, Bandar Bin Qasim$^3$, Daniel DeBra$^1$, and Robert Byer$^1$}

\address{
$^1$W.W. Hansen Experimental Physics Lab, Stanford University, Stanford, CA 94305 \\
$^2$NASA Ames Research Center, Moffett Field, CA 94035 \\
$^3$King Abdulaziz City for Science and Technology, Riyadh, Saudi Arabia 11442 \\
$^4$SN\&N Electronics, Inc. Santa Clara, CA 95050
}
\ead{saraf@stanford.edu}
\begin{abstract}

The UV-LED mission demonstrates the precise control of the potential of electrically isolated test masses that is essential for the operation of space accelerometers and drag-free sensors.  Accelerometers and drag-free sensors were and remain at the core of geodesy, aeronomy and precision navigation missions as well as gravitational science experiments and gravitational wave observatories. Charge management using photoelectrons generated by the 254nm UV line of Hg was first demonstrated on Gravity Probe B and is presently part of the LISA Pathfinder technology demonstration. The UV-LED mission and prior ground testing demonstrates that AlGaN UV‑LEDs operating at 255~nm are superior to Mercury vapor lamps because of their smaller size, lower power draw, higher dynamic range, and higher control authority. We show flight data from a small satellite mission on a Saudi Satellite that demonstrates “AC charge control” (UV-LEDs and bias are AC modulated with adjustable relative phase) between a spherical test mass and its housing. The result of the mission is to bring the UV-LED device Technology Readiness Level (TRL) to TRL-9 and the charge management system to TRL-7. We demonstrate the ability to control the test mass potential on an 89~mm diameter spherical test mass over a 20~mm gap in a drag-free system configuration. The test mass potential was measured with an ultra-high impedance contact probe. Finally, the key electrical and optical characteristics of the UV-LEDs showed less than 7.5\% change in performance after 12 months in orbit.

\end{abstract}

\maketitle

\section{Introduction}
A Modular Gravitational Reference Sensor (MGRS) (\cite{Sun2006},~\cite{Sun2009},~\cite{sun2011}) is being developed for space-based missions that require nanometer to picometer level precision test mass (TM) position measurement. The MGRS is a sensor for drag free control (\cite{lange1964thesis},~\cite{triad1974},~\cite{debra1999},~\cite{debra2011}) that uses optical sensing techniques to determine the spacecraft position relative to the TM, allowing the TM to float with near zero stiffness~\cite{Sun2006}. The spacecraft and the test mass housing blocks many disturbances including atmospheric drag, solar wind, radiation pressures, and mitigates thermal effects, (\cite{higuchi2009},~\cite{alfauwaz2011}) which otherwise may lead to the TM path deviating from the geodesic. However, highly energetic particles are still capable of penetrating through the spacecraft outer surface and charging the TM, either directly or via secondary electron emission (\cite{Wass2005},~\cite{Araujo2005}). This in turn leads to TM charging rates on the order of 5 to 200 positive charges per second~\cite{Sumner2009} for most flown or active experiments, depending on orbit and solar activity, and on the shielding provided by the spacecraft and the TM housing.  Breaking of contact between TM and its housing can leave the TM charged to as high as 0.5~V, depending on the materials coming in contact. Note that the TM to housing capacitance is of the order of a few to tens of pF for large gap instruments (\textgreater5~mm) and as much as 1~nF for GP-B (32~$\mu$m gap). Such charging leads to an electrostatic disturbance force that would corrupt the signal necessary for both the scientific measurement and drag-free control. Generally, the limit on the TM charge is 0.1~pC - 10~pC, requiring continuous or periodic activation of the charge management system.
 
Charge management is achieved through photoemission; missions such as Gravity Probe B~\cite{buchman1995} and LISA Pathfinder~\cite{Sumner2009} have used the 254~nm UV line of mercury lamps as the light source. Deep UV-LEDs operating at 255~nm have been identified as a new method to mitigate TM charging~\cite{Sun2006a},~\cite{Pollack2010}; newer UV-LED devices operating as low as 240~nm are also becoming commercially available~\cite{Olatunde2015}. Compared to Hg lamps, UV-LEDs are smaller and lighter, consume less power, have a wider spectrum selection, and a much higher dynamic range, with at least an order of magnitude improvement in each performance area. The power output is also very stable, with a lifetime of \textgreater 30,000~hours demonstrated in laboratory testing. As noted in~\cite{Weber2003}, applied fields can interact with DC biases or charge and lead to in-band force noise; by using UV-LEDs for AC charge transfer, charge management can be performed outside the science band. 

\begin{figure}[htb]
	\centerline{\includegraphics{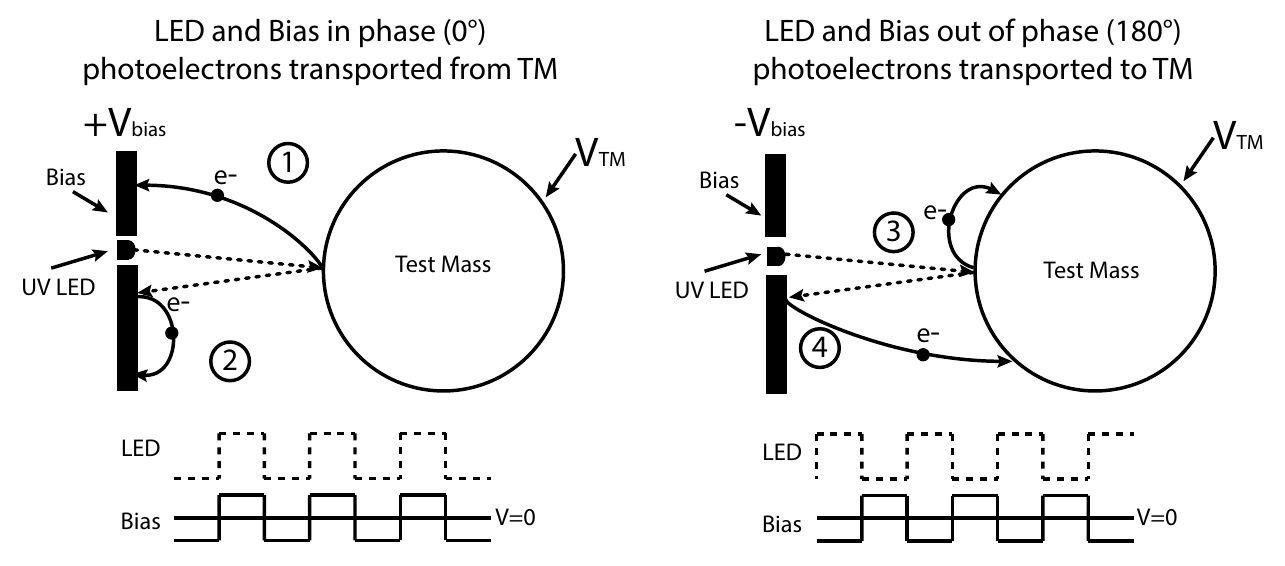}}
	\caption{Schematic showing the basic operation of AC charge management. The solid lines describe the electron path while the dashed lines describe the UV path. When the bias plate voltage relative to the housing (V$_{bias}$) is positive, (1) photoelectrons generated from the TM and (2) photoelectrons generated from the bias plate travel to the bias plate leading to an increase in the TM potential (V$_{TM}$). When V$_{bias}$ is negative, (3) photoelectrons generated from the TM and (4) photoelectrons generated from the bias plate travel to the TM leading to a decrease in V$_{TM}$.}
	\label{fig:cmsBasics}
\end{figure}

In this paper, we demonstrate an AC charge management system in a MGRS-like configuration using a single bias plate with large gap size, UV-LED light source, and a single large spherical TM as shown in Figure~\ref{fig:cmsBasics}.

AC charge management can be split into two cases: in-phase (positive charge transfer) or out-of-phase (negative charge transfer). In both cases, UV light is directed at the TM and photoelectrons are generated from its surface. Some of the UV light reflects back to the bias plate and photoelectrons are generated from the bias plate surface. In the positive transfer case, both drive signals are in phase so V$_{bias}$  is positive while the LED is turned on; generated photoelectrons are pulled towards the bias plate. When V$_{bias}$ is negative, the LED is off and no photoelectrons are generated. In the negative transfer case, because the drive signals are out of phase, V$_{bias}$ is negative when photoelectrons are generated. Thus, electrons are pushed from the bias towards the TM. When V$_{bias}$  is positive, the LED is off and no photoelectrons are generated. The rate of charging depends on UV optical power, coating properties such as workfunction, quantum efficiency, and reflectivity, and the surface roughness of the TM and bias plate.

\section{UV source properties and testing}

\begin{figure}[htb]
	\centerline{{\includegraphics{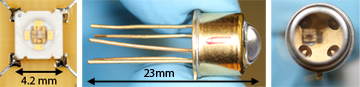}}}
	\caption{UV-LEDs of various packaging styles used in qualification level testing. From left to right: Surface Mount (SMD) affixed on a printed circuit board, Hemispherical (HS), and Tall Flat Window (TFW).}
	\label{fig:ledPackaging}
\end{figure}

The UV-LED source considered for MGRS charge management is an AlGaN based device supplied by Sensor Electronic Technology, Inc.~\cite{SET}. The UV-LEDs are available in several packages, such as those shown in Figure~\ref{fig:ledPackaging}, allowing them to be easily integrated inside a MGRS housing. The selected devices contain both an LED source as well as a witness photodiode, allowing for real-time monitoring of the UV output. Part numbers for the devices are UVCLEAN255-TO39HS/TO39TFW/SMD4.2FW, specially ordered with internal photodiodes. The output is centered at 255~nm with a Full Width at Half Maximum (FWHM) of 12~nm. Optical power output can be controlled from \textless1~nW to \textgreater100~$\mu$W by varying the forward current (compared to a dynamic range \textgreater140 for mercury lamps~\cite{Shaul2008}); having 5 orders of magnitude of control authority allows for the construction of a charge control scheme that is robust even when faced with high charging rates or changes in coating properties. Modulation greater than \textgreater10~kHz is also possible, allowing charge management to take place at frequencies far above the science band.
 
The effectiveness of the charge generation ultimately boils down to the energy of the photons output from the LED and the work function of the surfaces from which photoemission is required.  The photon wavelength (and energy) is determined by the LED material.  In the case of the chosen AlGaN UV-LEDs, the central wavelength of 255~nm translates to an average electron energy of 4.862~eV; a test mass coating with workfunction below this level will produce photoelectrons when exposed to the UV light.  For example, the work function of atomically clean Au thin film coatings in the 15-200~nm thickness range is nominally 4.9~eV, when exposed to air during normal handling procedures the work function can decrease down to 4.3~eV in a matter of days~\cite{Jiang1998}.  Similar changes in workfunction are also seen in the tested thin film coatings.  Because of the 12~nm FWHM, even an atomically clean Au thin film surface will produce photoelectrons.

We have performed extensive tests including lifetime, radiation, and MIL-STD-1540E~\cite{Perl2006} level thermal and vibration to validate device robustness for spaceflight. Lifetime tests in vacuum (currently \textgreater30,000 hours) and nitrogen (\textgreater12,000~hours) using hemispherical lens TO-39 devices, high fluence proton irradiation (63~MeV at a fluence of 2~protons/cm)~\cite{Sun2009a} using flat lens TO-39 devices, thermal vacuum (30 cycles at -34$^\circ$C to +71$^\circ$C), and vibration (14.07~g RMS for 3 minutes per axis)~\cite{Balakrishnan2011} using flat, ball, and hemispherical lens TO-39 and SMD devices have demonstrated the robustness of UV-LEDs.

\begin{figure}[htb]
	\centerline{\resizebox{!}{1.5in}{{\includegraphics{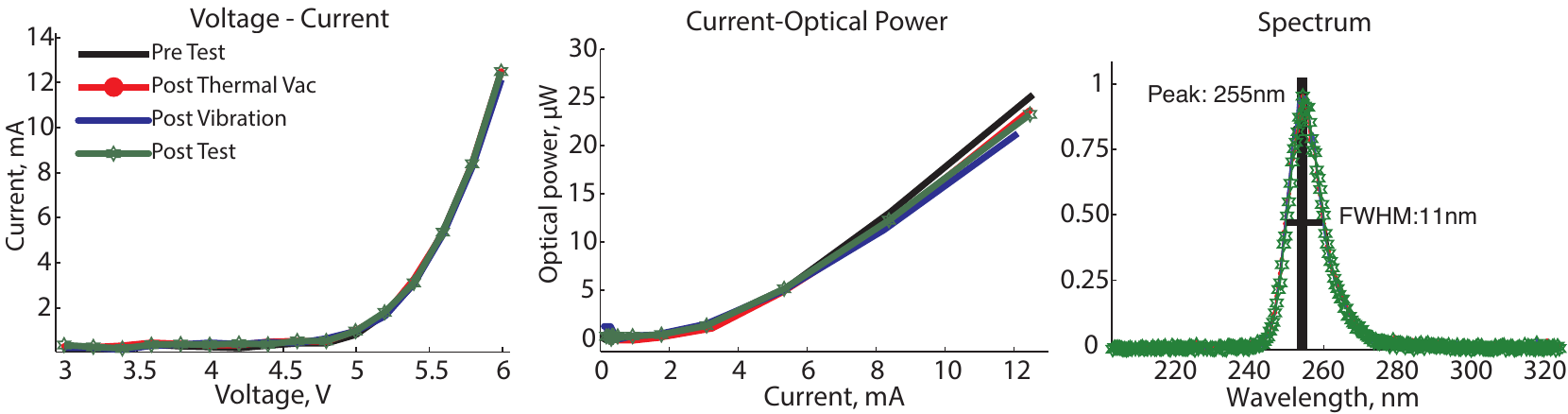}}}}
	\caption{Characteristic performance plots of an uncollimated flat window TO-39 packaged UV-LED taken during MIL-1540 level laboratory testing. Figures show data taking before testing, after thermal vacuum cycling, after shake, and after thermal cycling (post test). From left: Voltage (V) vs. Current (mA), Current (mA) vs. Optical Power ($\mu$W), and Spectrum.}
	\label{fig:ledCharPerf}
\end{figure}

Representative results from the MIL-1540 laboratory testing are shown in Figure~\ref{fig:ledCharPerf}. The IV (current versus voltage) curve was generated using an Agilent E3631A power supply, optical power was measured with a Newport 1931-C power meter with a 918D-UV OD3 detector head, and spectrum was measured with an OceanOptics MayaPro spectrometer with the LED driven at 10~mA. The 918D-UV has a 2\% uncertainty and ±2\% uniformity over the 220-349~nm range~\cite{Newport2010}, within which the emitted UV-LED spectrum falls. All properties were measured with DC current.

The IV curve for a diode is the fundamental measure of its PN junction characteristics and a family of well behaved IV curves is a good indication of diode chipset quality. During MIL-1540 level testing, the semiconductor chipset performance remains stable through all thermal and vibration tests, indicating that the electrical properties of the chipset are effectively unchanged. In the 5-10~mA range where most charge management will be performed, there is an increase in potential of less than 3\% in the tested diodes. Similarly, there is no shift in spectral peak or FWHM. This level of robustness indicates that UV-LEDs are a suitable candidate for spacecraft charge mitigation in applications currently using mercury lamps.

\section{AC (Active) charge management experiment}
\subsection{Experimental Setup}

\begin{figure}[htb]
	\centerline{\resizebox{!}{2.5in}{\includegraphics{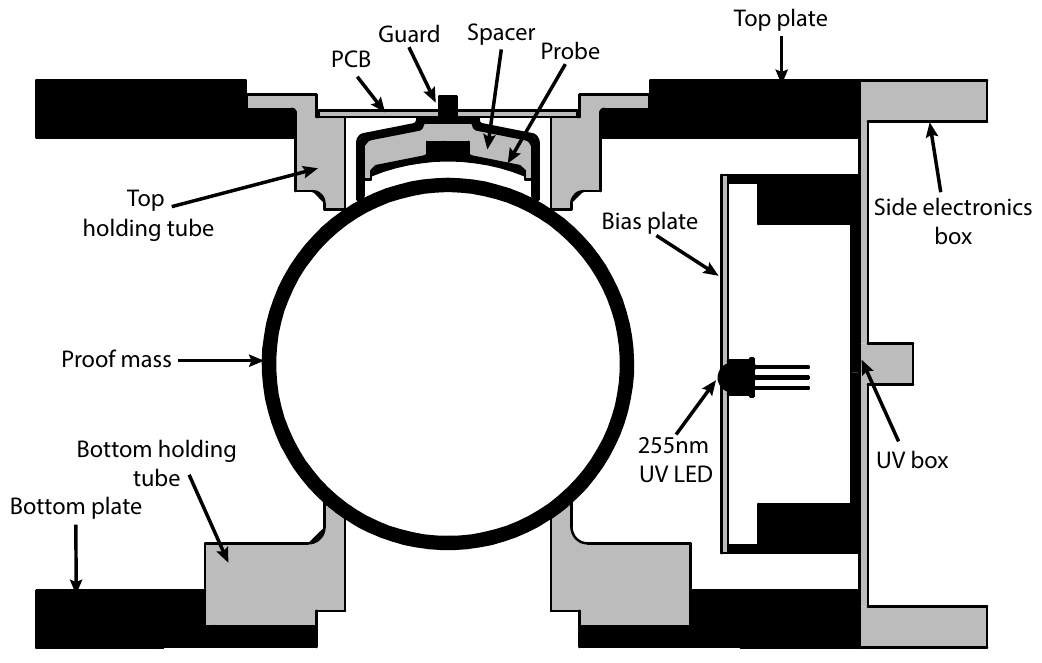}}}
	\caption{Cutaway schematic of AC charge management experiment structure. Total TM capacitance to ground is 17 pF.}
	\label{fig:cmsLabCutaway}
\end{figure}

\begin{figure}[htb]
	\centerline{\resizebox{!}{2.5in}{\includegraphics{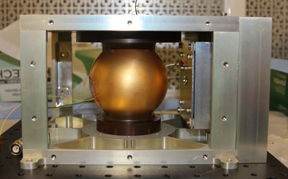}}}
	\caption{Prototype research system showing single bias plate, coated sphere, and Ultem holding tubes containing floating probes. The wire attached to the TM was used during early tests to calibrate the TM potential with the potential measured by the floating probe, but was removed during actual experimental runs.}
	\label{fig:cmsProtoSystem}
\end{figure}

The laboratory demonstration was performed using commercially available test equipment. A single 255~nm flat window TO-39 package was used as the UV source. A dual channel function generator (Agilent 33522A) modulated both the LED and Bias with a square wave at 100Hz with 50\% duty cycle, allowing precise control of relative phase. TM potential was read back using an electrometer (Keithley 6514) in voltage mode using guarded inputs. The integrated photodiode response was measured with a picoammeter (Keithley 6485) and was used for LED status monitoring.
A schematic showing the experimental setup is shown in Figure~\ref{fig:cmsLabCutaway}, and a photograph of the experimental structure is shown in Figure~\ref{fig:cmsProtoSystem}. The TM is a 88.9~mm diameter hollow Aluminum 6061-T6 sphere with 3.175~mm wall thickness. The exterior of the sphere was cleaned via an HF etch. A 20~nm Ti sticking layer was coated over the entire surface followed by a 150~nm Au layer, both via e-beam evaporation. The Ti improves Au adhesion and prevents alloying between the Al and Au layers. The TM is supported by two insulated Ultem-1000 holding tubes.

Bias is provided via a single Aluminum 6061-T6 square plate measuring 92~mm on an edge, located 20~mm away from the closest point of the TM. The bias plate is unpolished (standard extruded roughness of 3.5-7.5~$\mu$m)~\cite{mcmaster} and coated via e-beam evaporation with a 20~nm Ti sticking layer and~150 nm Au photoemission layer.

TM potential is measured using a single 38~mm diameter aluminum probe disk located 4~mm away from the sphere surface and housed inside a guard shell. The probe is curved so that its surface is always parallel to the TM. As a result, the magnitude of the electric field induced by the TM potential is equal along most of the probe. The probe is connected to the Keithley 6514 which returns the potential induced at the probe surface. The guard is driven to the probe potential by the Keithley 6514's built in voltage follower, helping shield the measurement from interference by driving electronics.

\subsection{Results}

\begin{figure}[htb]
	\centerline{\resizebox{!}{2in}{\includegraphics{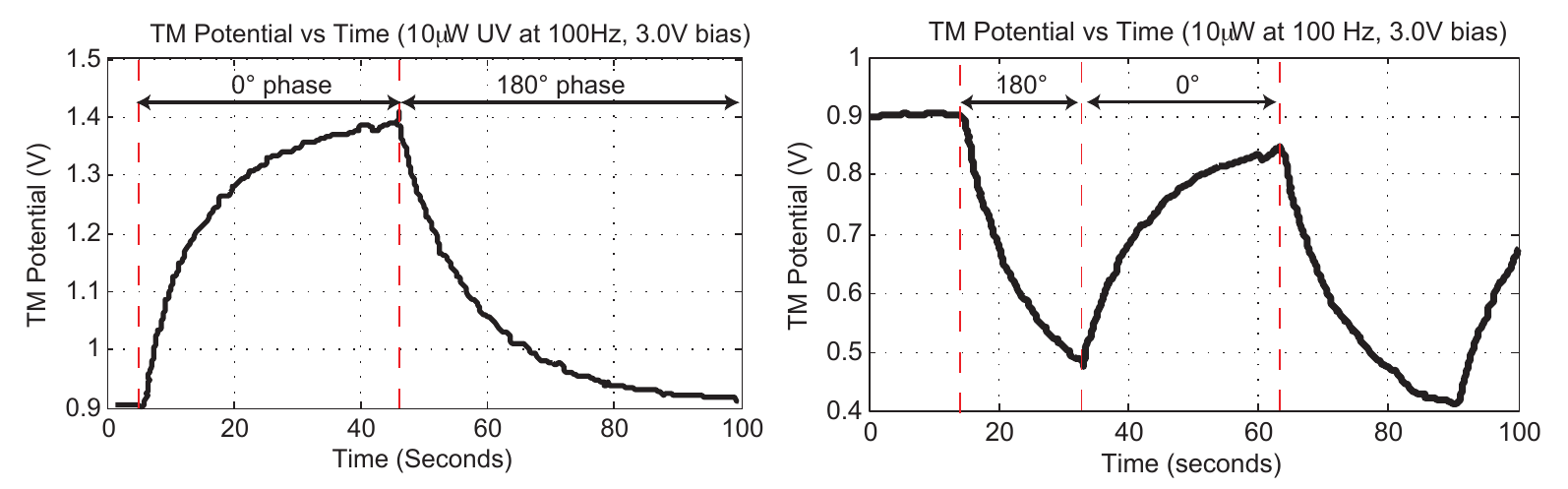}}}
	\caption{TM potential measured using a floating probe during AC charge management operation. Potential is measured relative to the system housing which is held at ground. Incident peak UV power is 10~$\mu$W modulated at 100~Hz square wave with 50\% duty cycle, with a 3.0~Vpp bias. TM capacitance to ground is 17~pF. Peak charging rates are 0.53~pA (positive) and 0.40~pA (negative).}
	\label{fig:labresults}
\end{figure}

During the experimental run shown in Figure~\ref{fig:labresults}, a single flat window TO-39 package UV-LED was modulated with a 100~Hz square wave with 50\% duty cycle and driven at 6.4~mA peak current, resulting in a peak optical power output of 10~$\mu$W. The bias was also modulated at 100~Hz, 3.0 ~V$_{pp}$. The initial TM potential is 0.9~V, due to the residual charge left on the sphere when disconnecting it from ground. On the left figure, both LED and bias were initially in phase with each other (0$^\circ$) leading to a peak positive charge transfer at a rate of 0.21~pA; upon reaching a peak, the bias phase was flipped to 180$^\circ$ leading to a peak negative charge transfer at a rate of 0.15~pA. The LED was then turned off for several minutes.   As shown in the right figure, the LED was then turned on with bias at 180$^\circ$ phase, leading to a peak negative charge transfer rate of 0.53~pA, and flipping the bias phase to 0$^\circ$ led to a peak positive charging rate of 0.40~pA.
%

The upper bound on maximum predicted charging rate under these conditions is 0.6~pA. Note that positive charging occurs more quickly (40~seconds) than negative charging (55~seconds). The rate at which the TM potential changes can be reduced by decreasing the LED optical power or decreasing the duty cycle during which the LED is switched on. Previous tests have shown that the UV-LED is capable of performing charge management when being driven at 10~kHz~\cite{Sun2006a}. Thus, an LED running in its normal operational range can out 1~nW average power pulsed for a single minimum-width cycle can emit as few as 6.4 photons. The high dynamic range of the UV-LEDs will allow for lower charge and discharge rates as required by actual system performance in orbit.

\section{The UV-LED Mission}
\subsection{Payload Design}
\begin{figure}[htb]
	\centerline{\resizebox{!}{2.5in}{\includegraphics{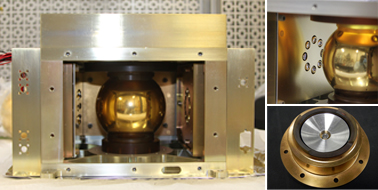}}}
	\caption{Flight model of UV-LED payload developed at Stanford. Clockwise from left: view of payload, 8 UV-LEDs directed towards the TM through openings in the gold-coated bias plates, charge amp and contact probe installed in gold coated Ultem holding tube.}
	\label{fig:uvledFlightModel}
\end{figure}

\begin{figure}[htb]
	\centerline{\resizebox{!}{2.5in}{\includegraphics{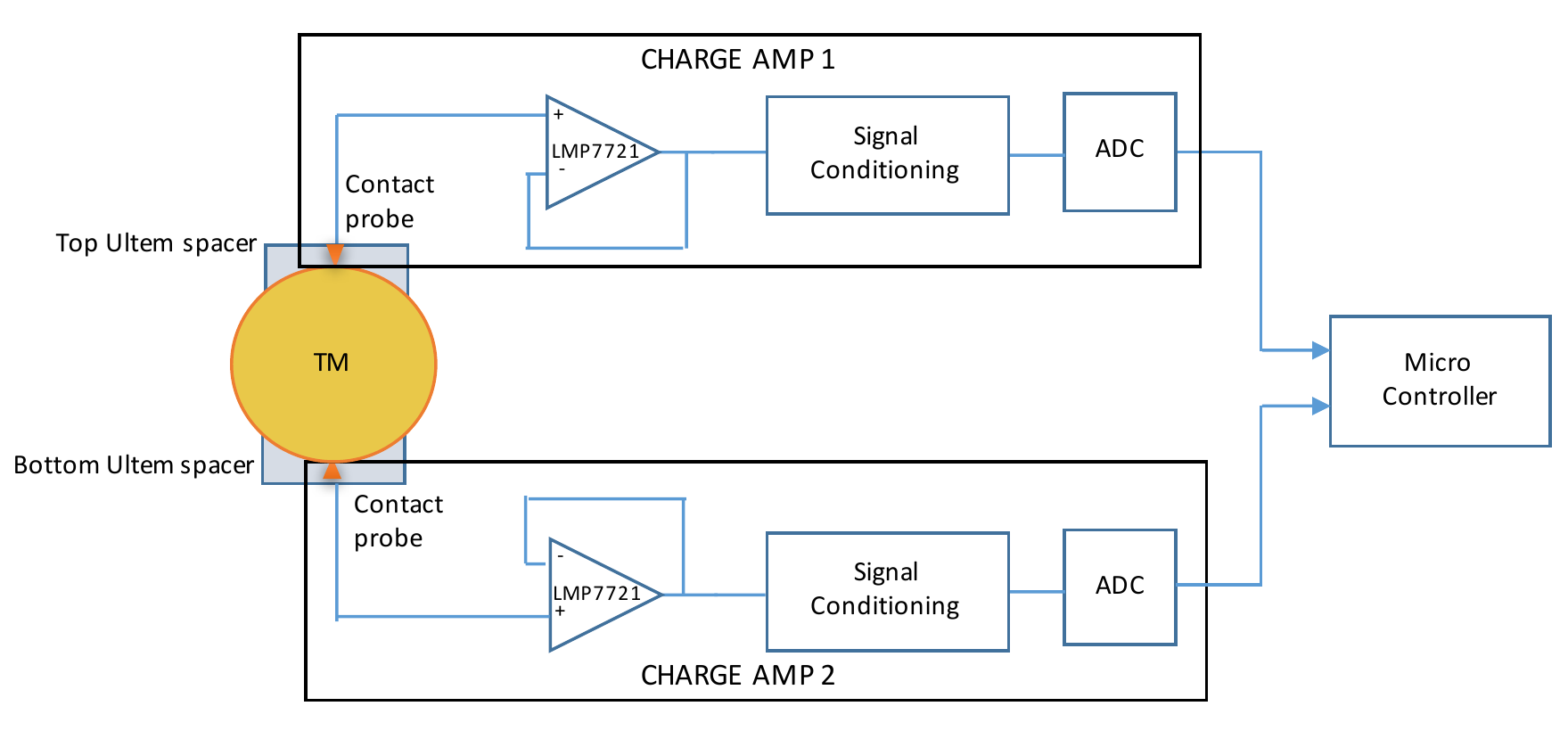}}}
	\caption{Charge amplifier schematic. The top and bottom charge amplifiers are independent TM charge measurement or equivalently potential measurement systems referred to as experiment 1 and experiment 2. The LMP7721 is an ultra-low input bias current operational amplifier configured as a high-impedance buffer to measure the TM potential. The gold coated Ultem holding tubes provide the required electrical isolation between the TM and the chassis as described earlier.}
	\label{fig:chargeAmpSchematic}
\end{figure}

Several modifications have been made at Stanford University to the design of the laboratory model  to reduce the size and improve the reliability and robustness of the payload for flight. Figure~\ref{fig:uvledFlightModel} shows a photograph of the UV-LED flight payload. The key modifications are listed below:

\begin{itemize}
\item The number of LEDs has been increased from 1 to 16, allowing for study and characterization of lifetime and performance statistics. This increase also allows a higher optical power to be directed at the TM, increasing the possible charging and discharging rate.
\item The experimental setup has been split into 2 identical subsystems called experiment 1 and experiment 2. Each subsystem has 8 UV-LEDs and a dedicated contact probe and charge amplifier circuit.  The signal processing chain is also duplicated. This redundancy of the measurement apparatus allows for the failure of a subsystem without compromising the mission. Figure~\ref{fig:chargeAmpSchematic} shows a schematic of the dual charge amplifier measurement system.
\item Both surface mount and flat window style LEDs are present, to test the effects of the space environment on different packaging and lens materials. 
\item The floating probe has been replaced with a contact probe and a charge amplifier with an input impedence of 10~$\Omega$ due to payload volume constraints. 
\item The number of bias plates has been increased from 1 to 4 for symmetry. All bias plates are gold coated and increased in size to minimize stray reflections. 
\item The top and bottom aluminum plates of the housing are grounded and not gold-coated. Therefore, photoemission from incident UV light will not occur from these two plates. However, electrically these plates can act as sinks for the electronic photocurrent from the bias plates and the TM. This is explained later in the paper.
\item The TM has been polished to a mirror finish before coating, to more closely represent the surface finish of a MGRS TM. 
\item Both Ultem holding tubes have been gold coated (with the exception of a small pad that contacts the TM). This coating helps prevent the buildup of electrical charge on the holding tubes. Additionally, it helps shield the sensitive contact probe electronics from interference caused by surrounding electronic boards.
\end{itemize}

\subsection{Satellite and Orbit Parameters}

\begin{figure}[htb]
	\centerline{\resizebox{!}{2.5in}{\includegraphics{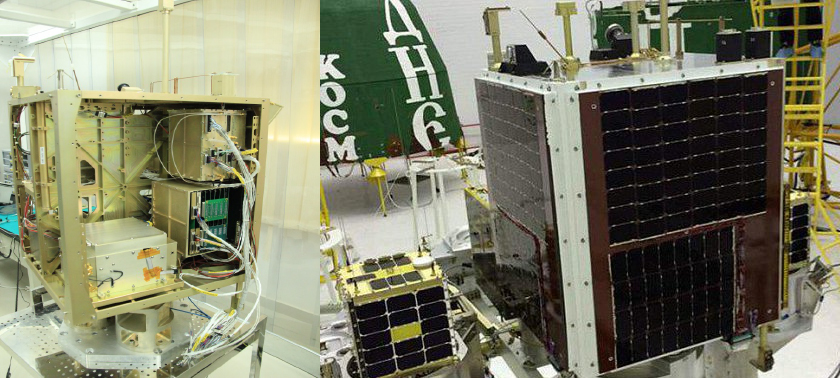}}}
	\caption{Left: SaudiSat-4 with the UV-LED payload integrated into the lower left corner undergoing final tests at KACST. Right: SaudiSat-4 ready for launch on a Russian Dnepr-1 rocket.}
	\label{fig:saudiSatPayload}
\end{figure}

A spacecraft demonstration of UV-LEDs and UV-LED charge management based on the research done at Stanford University (SU) was developed jointly by King Abdulaziz City for Science and Technology in Riyadh, Saudi Arabia (KACST) and NASA Ames Research Center (ARC). The goal of the mission is to bring the UV-LED device to TRL-9 and the charge management system to TRL-7 (available from~\cite{Engineering2011}). The UV-LED payload prototype was developed and tested at SU while the UV-LED  engineering (flight back-up) unit and flight unit (18~kg, 45~W) were constructed, functionally tested, and environmentally qualified at ARC as a collaboration of the ARC and SU teams. The SaudiSat-4 satellite shown in Figure~\ref{fig:saudiSatPayload} was developed and built by KACST, where the payload integration with the satellite along with the functional and environmental testing of the entire hardware and software took place. This was primarily a KACST, ARC collaboration. The UV-LED experiment was launched on June 19$^{th}$ 2014 on a Dnepr-1 rocket from Plesetsk cosmodrome in Russia as one of the 33 satellites placed in orbit~\cite{kosmotras2014}.  The Saudi-Sat 4 (110~kg total mass, 61~W to 86~W power) was placed in a 97.6~minute period, 7,023~km semi-major axis, Sun-synchronous orbit. Contact with the spacecraft was established as soon as practical and science data acquisition started in December 2014 and continued to January 2016. Mission lifetime was expected to be at least one month followed by periodic payload turn-on and measurements. The ability for the UV-LEDs to mitigate actual space-based charging and the effects of radiation on the UV-LED device performance were studied. Extensive studies of diode parameters, and charge management efficiency were performed on the pre-flight instrument and repeated after 12 months in orbit. The test of the optical and electrical parameters of the diodes in a space environment allows us to estimate the performance of a drag-free sensor in a highly demanding and extended mission like LISA where the test mass dynamics allows a total of 3.2~pC total charge imbalance~\cite{Sumner2009}.

\section{Flight data}

A set of science scripts were prepared at Stanford that allowed measurements of the LED electrical and optical characteristics. Additional scripts were designed to exercise charge management of the TM by raising and lowering the potential (or equivalently the charge) of the sphere in a controlled manner using UV-LED light. These scripts were uploaded to the spacecraft via a ground station in Saudi Arabia and the results of the experiments were stored in the spacecraft bus memory. The results were then downloaded to the ground station when contact with the spacecraft was established. 

\subsection{UV-LED Characteristics}
\begin{figure}[htb]
	\centerline{\resizebox{!}{2.5in}{\includegraphics{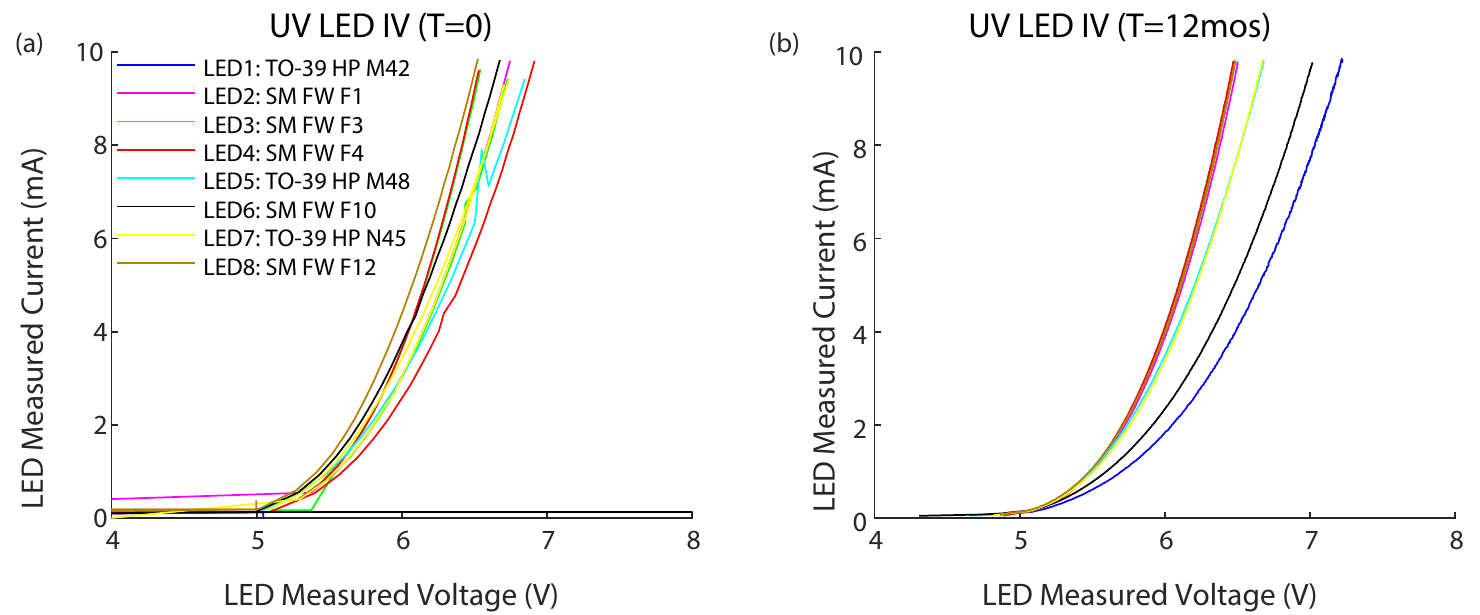}}}
	\caption{Comparison of IV (current-voltage) curve generated during a VIP (Voltage, Current, Power) test at (a) payload turn on, and (b) after 12 months in orbit. The legend defines the package, lens, and LED ID.  The package is either TO-39 or surface mount (SM); the lens is either flat window (FW) or hemispherical (HP).}
	\label{fig:compareIV}
\end{figure}

\begin{figure}[htb]
	\centerline{\resizebox{!}{2.25in}{\includegraphics{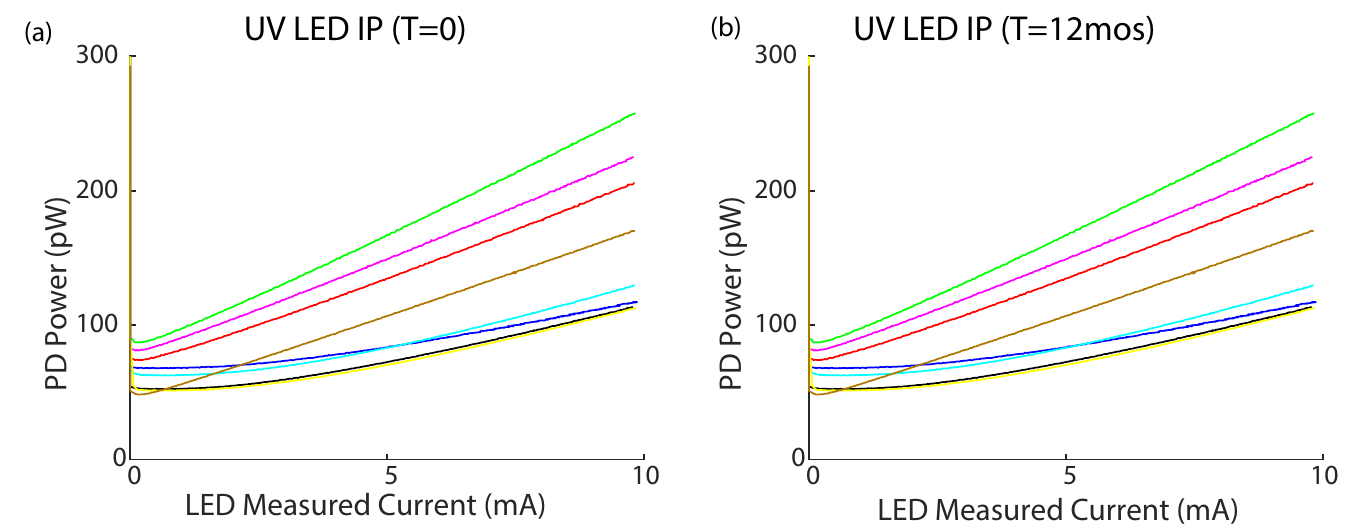}}}
	\caption{Comparison of IP (current-power) curve generated during a VIP (Voltage, Current, Power) test at (a) payload turn on, and (b) after 12 months in orbit, using witness photodiode power as the measure.  The power output of the LEDs did not show any significant change after 12 months in orbit.  Legend matches Figure~\ref{fig:compareIV}.}
	\label{fig:compareIP}
\end{figure}

\begin{figure}[htb]
	\centerline{\resizebox{!}{2.25in}{\includegraphics{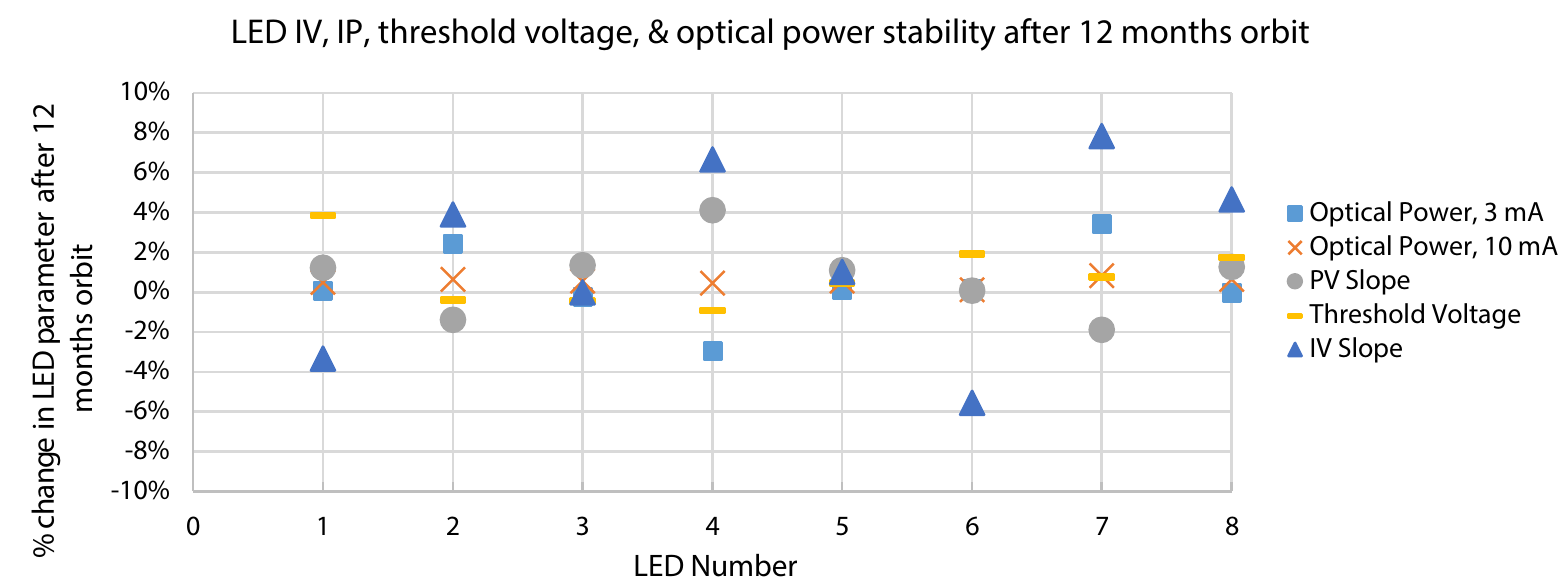}}}
	\caption{Change in values of key UV-LED performance parameters after 12 months in orbit.}
	\label{fig:keyValueChange}
\end{figure}

The current-voltage (IV) and power-current (PV) characteristics of the UV-LEDs were measured after the payload was turned on. The threshold voltage was approximately 5.5~V for the AlGaN diodes, matching data from individual LED qualification and ground tests of the payload. Figure~\ref{fig:compareIV}(a) shows the IV curve at turn on, and Figure~\ref{fig:compareIP}(a) shows the PV curves at payload turn-on. The threshold voltage and slope of the IV characteristics measured in space matched the ground measurements to within 5\%.  These measurements were repeated after 12 months in orbit. There was no significant change in the performance of the LEDs. Figure~\ref{fig:compareIV}(b) and Figure~\ref{fig:compareIP}(b) shows the results from the same tests after 12 months in orbit. Figure~\ref{fig:keyValueChange} shows the percentage change in the key properties, including optical power, IV slope, PV slope, and threshold voltage after 12 months in orbit.  The vast majority of LEDs exhibit less than 5\% change, and the properties of all LEDs show change of less than 7.5\% after aging on orbit.

\subsection{AC charge management}
Several experiments on AC charge management were conducted by varying the experimental parameters listed in Table~\ref{tab:programmableParameters}.

\begin{table}[htb]
\centering
\caption{Programmable parameters on the UV-LED experiment and the corresponding range of values exercised.}
\label{tab:programmableParameters}
\begin{tabular}{c|c}
\textbf{Parameter}          & \textbf{Range of Values} \\ \hline
Bias Plate Baseline Voltage & -2.5 to +2.5 V             \\
Bias Plate Offset Voltage   & -2.5 to +2.5 V             \\
UV-LED Phase                & 0-360$^\circ$            \\
Number of UV-LEDs           & 1 to 8 per experiment        \\
UV-LED Current              & 0 to 10 mA                \\
UV-LED Duty Cycle           & 0 to 100\%                 
\end{tabular}
\end{table}
 
\begin{figure}[htb]
	\centerline{\resizebox{!}{1.5in}{{\includegraphics{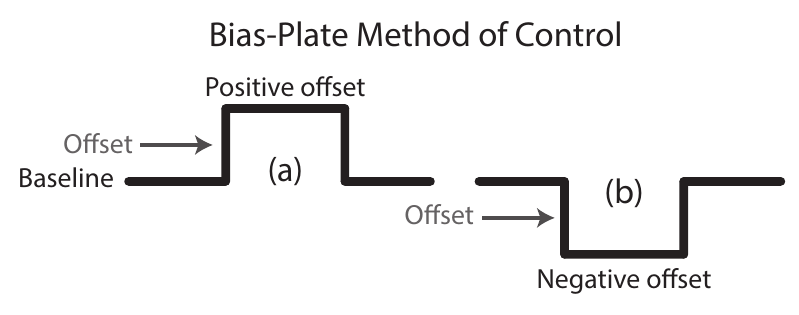}}}}
	\caption{Square wave of the bias plate voltage in terms of the bias plate baseline voltage and positive offset voltage (a) and negative offset voltage (b).}
	\label{fig:sqWave}
\end{figure}

The sphere potential is controlled by the voltage on the bias plate that are set by a combination of DC baseline voltage and AC offset voltage as shown in Figure~\ref{fig:sqWave}. 

\begin{figure}[htb]
	\centerline{\resizebox{!}{2.5in}{{\includegraphics{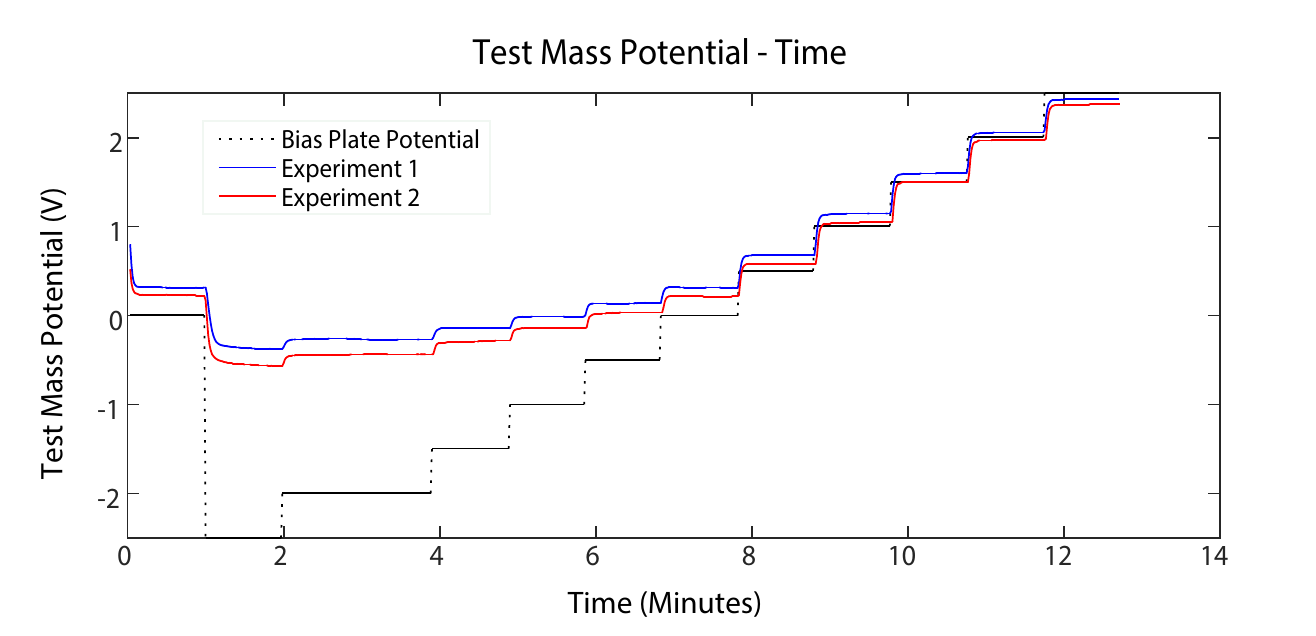}}}}
	\caption{Test mass potential as the bias plate offset is raised from -2.5~V to +2.5~V in steps of 0.5~V. Data is shown for both experiments. AC charge control phase was 0$^\circ$ with a duty cycle of 40\%. All 8 LEDs were driven at a current of 10~mA}
	\label{fig:measPotential}
\end{figure}

The initial set of experiments were conducted on both individual UV-LED set-of-eight banks (referred to as Experiment 1 and Experiment 2) to determine the performance of the system. As shown in Figure~\ref{fig:measPotential}, the UV-LED bank of Experiment 2 shows a slight negative bias compared to Experiment 1, caused by an electronic offset in the charge measurement system.

\begin{figure}[htb]
	\centerline{\resizebox{!}{2.5in}{{\includegraphics{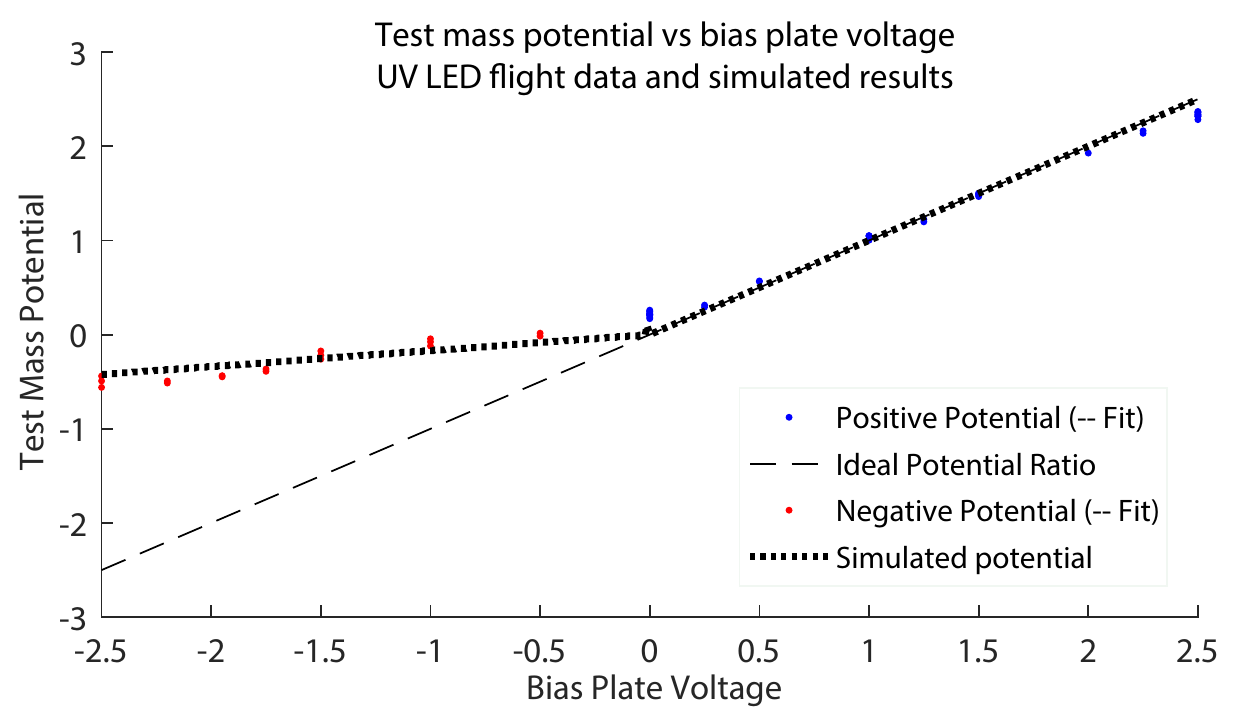}}}}
	\caption{Bias Plate Potential vs TM Potential data from flight data and simulation. The ideal ratio would be 1:1 between the bias plate and TM  potential.  Details of the simulation are given in Section~\ref{sec:EfieldSim}.}
	\label{fig:vTMvBias}
\end{figure}

The TM potential showed a linear variation with the bias plate potential as shown in Figure~\ref{fig:vTMvBias}. For positive bias plate potentials the slope is close to +1, as expected. However, for negative potentials the TM the slope is about -0.2. This lower slope is caused by the connection of the top and bottom plates of the housing to ground. This ground connection provides a sink for photoelectrons emitted from the TM and the housing. The electric field configuration in the housing is consequently modified from the case where all six sides of the housing are photoemissive and driven by a common V$_{bias}$. This will be explained in Section~\ref{sec:equivCircuit} with the help of a simplified system model, and in Section~\ref{sec:EfieldSim} through the use of simulation. 

\begin{figure}[htb]
	\centerline{\resizebox{!}{2.5in}{{\includegraphics{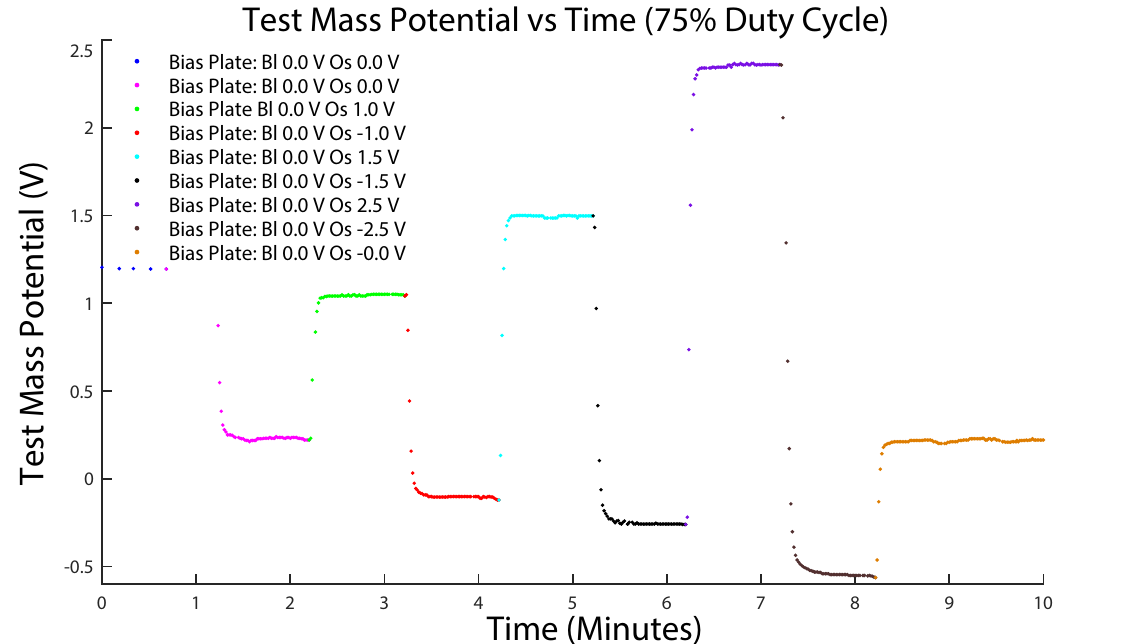}}}}
	\caption{Charge control experiment with UV-LEDs at 75\% duty cycle with fixed baseline (Bl = 0~V) and varying Offset (Os) voltages of the bias plates to obtain the corresponding TM potentials. Response time is a few seconds.}
	\label{fig:tmResponseFlight}
\end{figure}

Figure~\ref{fig:tmResponseFlight} shows charge control with the LEDs operating at 75\% duty cycle and a programmed combination of baseline and offset. This combination yields a net bias plate potential that cycles between positive and negative voltages. The test mass potential stabilizes rapidly within a second to the commanded bias potential. Analysis of the response time for a LED duty cycle between 40\% and 90\% yields a TM potential d$V_{TM}$/d$t$ in the range of 0.4~V/sec to about 1~V/sec. The measured TM capacitance to ground C2 is 28~pF as shown in Table~\ref{tab:measuredCaps} later in the paper.  Therefore, the TM charging rate d$Q_{TM}$/d$t$ = C2 d$V_{TM}$/d$t$ is calculated to be between 11.2~pA to 28~pA for 40\% and 9\% duty cycles respectively. This is equivalent to a maximum charge transfer rate of 1.75$\times$10$^8$~e$^-$/sec. In a well-designed drag-free system, the expected worst case charging rate is estimated under 1000~e$^-$/sec. Therefore, the UV photon flux could be lowered by 5 orders of magnitude and still meet the charge control requirements of the MGRS.
 
\subsection{TM, housing, and charge amplifier equivalent circuit}
\label{sec:equivCircuit}
\begin{figure}[htb]
	\centerline{\resizebox{!}{2.5in}{{\includegraphics{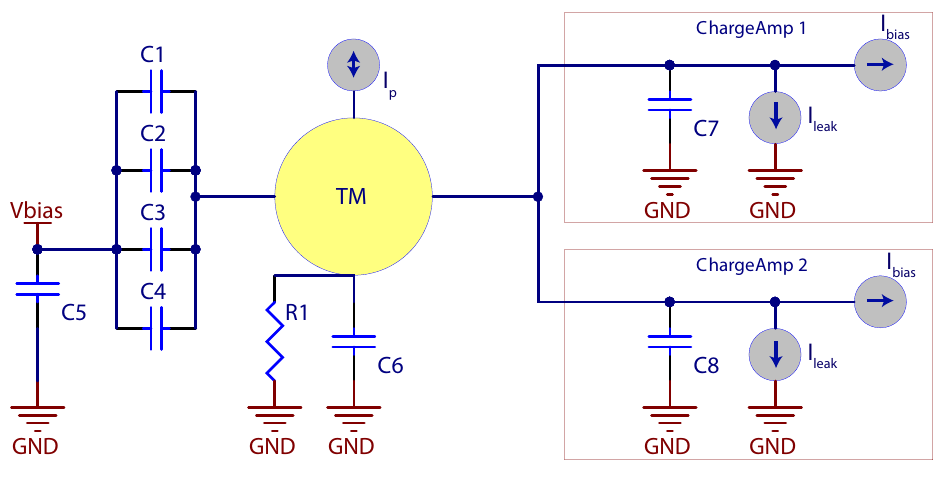}}}}
	\caption{The model circuit of the TM and charge management system in the UV-LED experiment showing the coupling, parasitic capacitances and current sources. I$_p$ is the electronic photocurrent that is sourced or sunk by the TM based on the potential of the bias plates.}
	\label{fig:modelCircuit}
\end{figure}

The equivalent circuit of the test mass mounted on an Ultem spacer inside the housing and connected to the two sets of charge amplifiers is shown in Figure~\ref{fig:modelCircuit}. The four gold-coated photoemissive bias plates that form the sides of the housing are electrically connected. C1, C2, C3 and C4 are the modeled capacitances between the TM and the four electrodes. Additionally, the top and bottom plates of the housing are connected to ground and the capacitance from the TM to these plates is modeled as C6. C5 is the capacitance between the bias plates and the two ground plates. R1 is the leakage resistance of the Ultem spacer and is of the order of 10$^{14}\Omega$. C7 and C8 are the front-end amplifier capacitances to ground. Table~\ref{tab:measuredCaps} tabulates the measured capacitances using an Agilent 4263B LCR meter.  The electronic photocurrent is modeled as a current source. In the case of positive bias plate potentials, the photoelectric current will flow from the TM to the plate, thus raising the TM potential. In the case of negative plate potentials, the photoelectric current will flow from the housing to the TM thus lowering the TM potential.  Table~\ref{tab:measuredCaps} tabulates the measured capacitances using an Agilent 4263B LCR meter.  The front-end amplifier consists of a contact probe to sense the TM potential followed by a high-impedance buffer circuit built around an ultra-low bias current operational amplifier LMP7721~\cite{Instruments2014}. The LMP7721 has a leakage current of 3~fA. Careful circuit design and guard ring techniques result in the FR4 printed circuit board (PCB) leakage to be of the order of a few~pA. The output of the voltage follower is signal-conditioned using active filters and gain stages followed by analog to digital conversion (ADC). The ADC output is the data stream presented to the microcontroller for digital processing as shown in Figure~\ref{fig:chargeAmpSchematic}. The charge amplifiers are modeled as a bias current I$_{bias}$ from the front-end amplifier and a leakage current I$_{leak}$ to ground through the PCB and operational amplifier.  This leakage current is of the order of a few~pA. 

\begin{figure}[htb]
	\centerline{\resizebox{!}{2in}{{\includegraphics{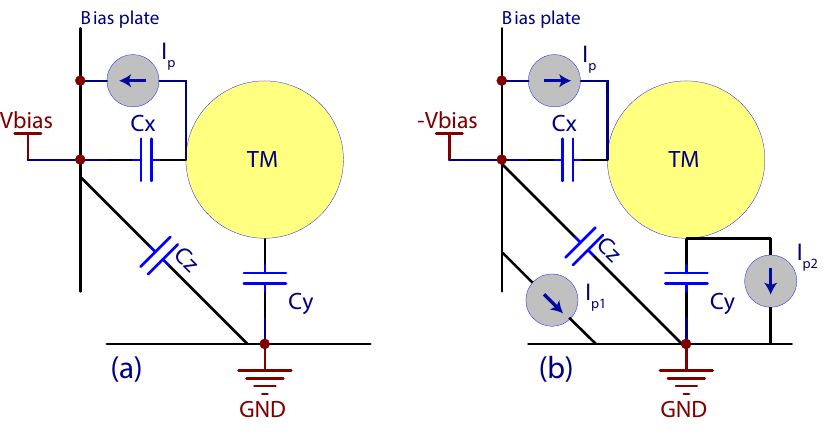}}}}
	\caption{Simplified model of the charge management system. Cx, Cy and Cz are capacitances measured in Table~\ref{tab:measuredCaps}. I$_p$ is the photoelectric current that flows from the bias plates to the TM. a) The bias plates are at a positive potential relative to ground. I$_p$ flows from the TM to the bias plates thus raising the potential of the TM. b) The bias plates are at a negative potential. I$_p$ flows from the bias plates to the TM thus lowering the potential of the TM. Additionally, I$_{p1}$ and I$_{p2}$ flow from the bias plates and the TM to the ground plates of the housing, respectively.}
	\label{fig:CMSystemCapModel}
\end{figure}

Two simplified models have been developed in order to understand and explain the data from Figure~\ref{fig:vTMvBias} for positive and negative bias plate potentials as shown in Figure~\ref{fig:CMSystemCapModel}. In the case of positive bias plate potentials modeled in Figure~\ref{fig:CMSystemCapModel}(a), photoelectric current I$_p$ flows from the TM to the bias plates since the plates are at the highest positive potential in the system. The TM potential will consequently rise until it approximately equals the bias plate potential. This explains the fitted slope of approximately 1 in the payload data shown in Figure~\ref{fig:vTMvBias}. The photoelectric current will settle at a value that will compensate for charge leakage from the TM due to the contact probe, PCB and the front-end amplifier. In the case of a free-floating TM in a drag-free configuration, this current should become zero.

In the model shown in Figure~\ref{fig:CMSystemCapModel}(b), the bias plates are shown connected to a negative potential. photoelectric current I$_p$ will now flow from the bias plates to the TM. Additionally, photoemission from the TM will result in a photoelectric current I$_{p2}$ to the top and bottom housing plates since these grounded plates are at a higher potential than the TM. Finally, photoemission from the housing will cause an electronic current I$_{p1}$  from the negative bias plates to the ground plates. I$_{p1}$ will flow continuously since the potentials of the bias plates and the ground plates of the housing are fixed. The TM potential will now follow the bias plate potential based on the capacitive divider formed by Cx and Cy. In other words, for a charge Q on the TM, Q = Cx*( V$_{bias}$ - V$_{TM}$) = Cy*V$_{TM}$. Therefore the voltage ratio V$_{TM}$/V$_{bias}$ = Cx/(Cx+Cy). Using the capacitance measurements from Table~\ref{tab:measuredCaps}, this ratio works out to 0.18. This is close to the fitted slope of 0.23 for negative bias plate potentials from Figure~\ref{fig:vTMvBias}.

\begin{table}[htb]
\centering
\caption{Capacitance measurements between various elements of the charge management setup.}
\label{tab:measuredCaps}
\begin{tabular}{c|c}
\textbf{Element}                   & \textbf{Measurements on EM} \\ \hline
Electrode-to-TM Capacitance (Cx)     & 6.1~pF                                     \\
TM-to-Ground Capacitance (Cy)        & 28~pF                                      \\
Electrode-to-Ground Capacitance (Cz) & 44~pF                                     
\end{tabular}
\end{table}

The capacitance measurements were performed on the engineering model at Stanford that has slight differences from the flight model in its housing situated inside the satellite. Therefore, the electric field distributions and consequently the capacitances will not be the same and will result in variations of the measured TM potential between the laboratory model and the flight unit in orbit.  
Again, the photoelectric current I$_{p1}$ will be continuous, while photoelectric current I$_{p2}$ will settle to a value that will compensate for the charge leakage through the charge amplifier electronics from PCB leakage, bias current for the operational amplifier etc. It is important to note in the above discussion that “conventional” current flows in a direction that is opposite to the direction of the photoelectric current shown in the figures.

\section{Electric field simulation}
\label{sec:EfieldSim}

In order to validate the analysis of the charging and discharging physics in Section 5, a MATLAB based E-field simulation was developed, giving a better understanding of the interaction between the charged test mass, driven bias plates, and grounded payload housing.  The methods and setup used for the simulation are shown in Section~\ref{sec:eFieldProcess}

\begin{figure}[htb]
	\centerline{\resizebox{!}{2.5in}{\includegraphics{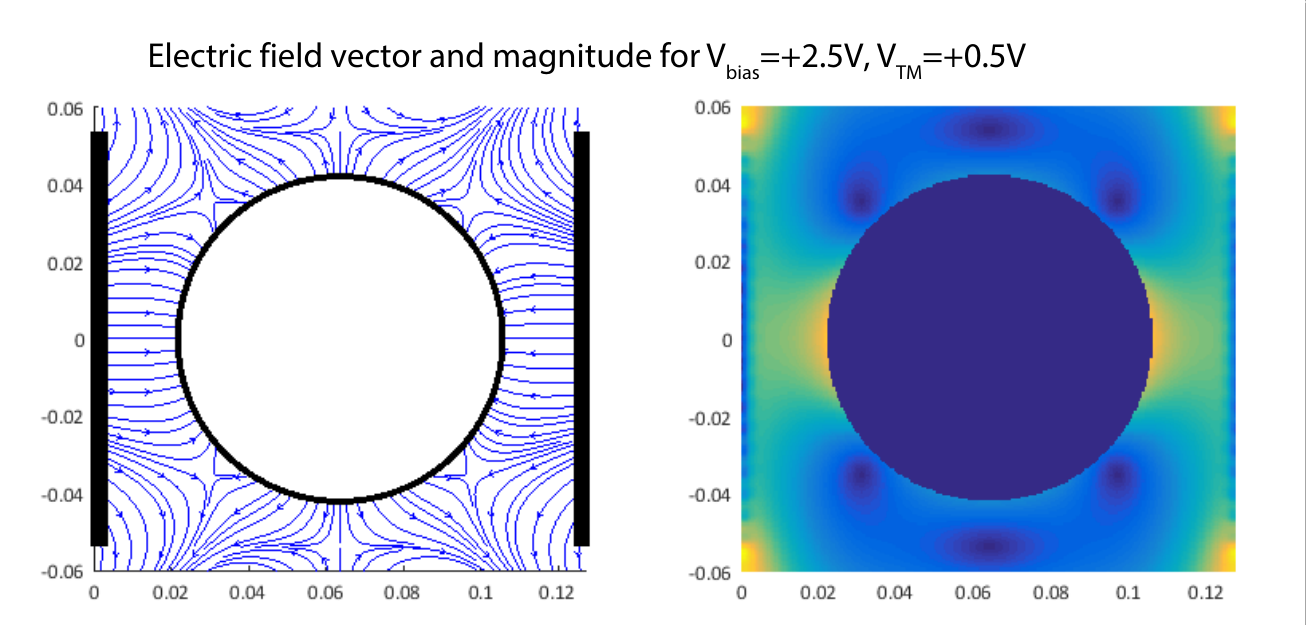}}}
	\caption{Electric field streamline and magnitude for V$_{bias}$ and V$_{TM}$ \textgreater 0.  Geometry units are in m.}
	\label{fig:eFieldSteamPositive}
\end{figure}

\begin{figure}[htb]
	\centerline{\resizebox{!}{2.5in}{\includegraphics{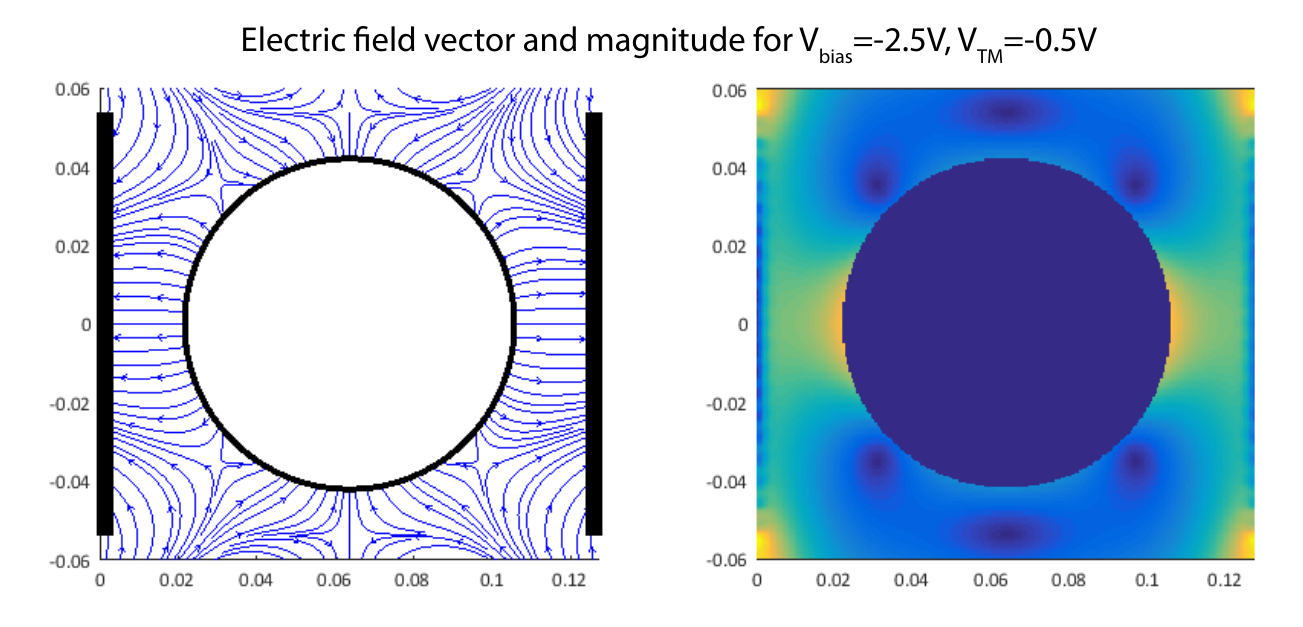}}}
	\caption{Electric field streamline and magnitude for V$_{bias}$ and V$_{TM}$ \textless 0.  Geometry units are in m.  Note that the field lines point in the opposite direction from Figure~\ref{fig:eFieldSteamPositive}}
	\label{fig:eFieldSteamNegative}
\end{figure}

The electric field streamlines and magnitudes are shown in Figure~\ref{fig:eFieldSteamPositive} for the case of V$_{bias}$ and V$_{TM}$ \textgreater 0 and Figure~\ref{fig:eFieldSteamNegative} for V$_{bias}$ and V$_{TM}$ \textless 0.  It is clear from the electric field lines that the electron path is greatly influenced by the presence of the grounded housing.  In both cases, the field lines connect the equator of the test mass and the center region of the bias plate.  However, the field lines from the test mass poles and edge region of the bias plates connect to the grounded housing.  Note from Figure~\ref{fig:cmsBasics} the requirements for positive and negative charging.  In the configuration shown in Figure~\ref{fig:eFieldSteamPositive}, electrons generated by the test mass are easily swept away to the bias plate; those generated by the bias plate are pushed back to the bias plate or swept to ground.  The electrons generated near the test mass poles are also swept back to the test mass.  Thus, the bias plate effectively controls the flow of electrons.  In the case of Figure~\ref{fig:eFieldSteamNegative} with V$_{bias}$ and V$_{TM}$ \textless 0, while electrons generated by the bias plate are easily swept to the test mass, those generated by the test mass near the poles are swept towards the housing, counteracting each other giving the bias plate less control authority.  

The maximum amplitude of V$_{TM}$ for a given V$_{bias}$ can be computed by taking the magnitude of E at the equator and poles, E$_{Eq}$ and E$_{Pole}$ of the test mass for a given V$_{bias}$.  The maximum amplitude of V$_{TM}$ is when E$_{Eq}$=E$_{Pole}$.  Once E$_{Pole}$ dominates, the electrons primarily flow towards the grounded housing and not the test mass.  Running the simulation across the range of V$_{bias}$ gives the results in Figure~\ref{fig:vTMvBias}.  The simulation closely matches the results from flight; when V$_{bias}$\textgreater0, the V$_{TM}$ closely tracks V$_{bias}$.  However, when V$_{bias}$\textless0, there is approximately a -1/5 slope, governed by the structure geometry and system capacitances.

\section{Conclusions}

The drag-free community has shown great interest in an LED source for charge management because of the significant advantages that it offers over traditional Hg lamp systems, especially in terms of system weight, volume, power, low thermal burden and operational flexibility. Laboratory research, ground measurements and flight data show that a charge management system based on low-power, high-bandwidth UV-LEDs can control test mass potential rapidly and with a high degree of fidelity. AC charge control allows us to exercise charge management without introducing noise in the MGRS signal detection band. We have demonstrated the ability to raise and lower the potential of the test mass using a combination of DC bias and a superimposed AC signal on the housing plates along with a phased AC drive of the UV-LEDs. The measured TM potential tracks the housing potential linearly with a high degree of accuracy. Detailed analysis of the flight data shows minimal degradation of the electrical and optical characteristics of the UV-LEDs, allowing the devices to be used on long-duration missions and in harsh space environments.
 
We continue to monitor the performance of the payload diodes with periodic measurements of the diode electrical and optical characteristics along with charge management reliability and repeatability. The increased variety of surface coatings provides a larger design space for the TM, allowing for better performance and higher survivability of the system. In the future, we will continue research on coating properties, AC and passive charge management techniques, fiber-optic UV light delivery systems and integration of the UV-LED charge management system into a complete MGRS with applications to new drag-free missions like LISA and GRACE follow-on.

\ack
The authors acknowledge the guidance and vision of Dr. Simon P. Worden, director of NASA Ames Research Center from 2006-2015, in promoting the concept of advancing space science and technology on small satellites under the aegis of international collaborations.

We are grateful to Dr. Turki Saud Bin Mohammed Al-Saud, president of KACST, for his leadership in the development of space science and technology at the KACST-Stanford joint center for excellence.  The work was funded by KACST and NASA and supported by the W.W. Hansen Experimental Physics Laboratory at Stanford University.

Karthik Balakrishnan was supported by the US Department of Defense and American Society for Engineering Education (ASEE) via the National Defense Science and Engineering Graduate (NDSEG) Fellowship.

\section*{Appendix A: Electric Field Simulation}
\label{sec:eFieldProcess}

\begin{figure}[htb]
	\centerline{\resizebox{!}{1.5in}{{\includegraphics{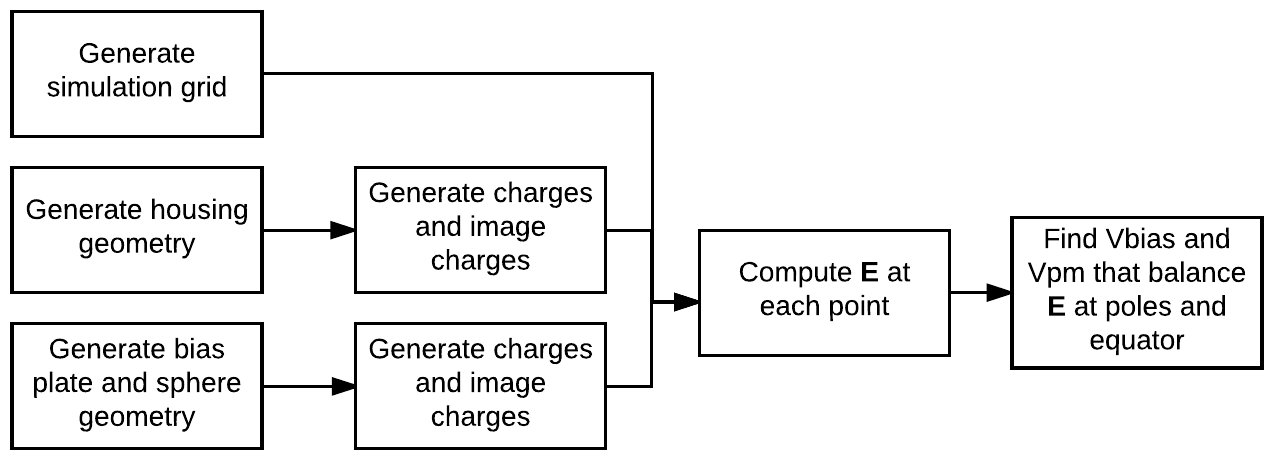}}}}
	\caption{Flowchart of E field simulation}
	\label{fig:eFieldSimFlow}
\end{figure}

Figure~\ref{fig:eFieldSimFlow} shows the high level process by which the electric field and VBias vs. V$_{TM}$ relationship is determined.  

First, the payload geometry is generated, including the housing, bias plates, and  test mass. Note that the grounded holding tube coating is not included in the simulation.  The geometric values used are shown in Table~\ref{tab:simGeo}.

\begin{table}[htb]
\centering
\caption{Dimensions of payload used for E field simulation}
\label{tab:simGeo}
\begin{tabular}{c|c}
\textbf{Geometric property} & \textbf{Value}    \\ \hline
Test mass radius            & 0.0425 m          \\
Test mass center            & (0.0637, 0, 0) m  \\
Housing height, width       & 0.1274 m, 0.136 m \\
Bias plate height, width    & 0.138 m, 0.106 m 
\end{tabular}
\end{table}

The test mass is modeled using discrete charges while the bias plate is modeled as a continuous surface.  Next, image charges are generated to ensure that the field lines are normal to each grounded and conductive surface.  Image charges about the grounded housing are generated by simply reflecting the base charges about the housing and negating the charge magnitude.  Image charges due to the conductive test mass, on the other hand, are generated based on the methods found in~\cite{Jackson1999}.  Finally, a mesh is generated inside the volume of the payload and E$_{x,y,z}$ at each point are computed.  For the bodies modeled as discrete charges, E is computed by summing up the electric field due to each point charge.  Simultaneously, the following expression is evaluated over the entire area of the bias plate to find its contribution to the electric field:

\begin{equation}
	\label{eq:basicPlateField}
    E = k_e \sigma \int \frac{\vv{r}-\vv{r}'}{|\vv{r}-\vv{r}'|^3}dA = k_e \sigma \int\int \frac{\vv{r}-\vv{r}'}{|\vv{r}-\vv{r}'|^3}dy'dz'
\end{equation}

where y and z are parallel to the bias plate surface, $\vv{r}$ is the vector from the center of the bias plate to the point in space where E is being computed, and $\vv{r}'$ is the vector from the center of the bias plate to any location on the bias plate.  
\section*{References}
\bibliographystyle{unsrt}

\bibliography{bibliography}

\begin{thebibliography}{10}

\bibitem{Sun2006}
Ke-Xun Sun, Graham Allen, Scott Williams, Saps Buchman, Dan DeBra, and Robert
  Byer.
\newblock {Modular Gravitational Reference Sensor: Simplified Architecture to
  future LISA and BBO}.
\newblock {\em Journal of Physics: Conference Series}, 32:137--146, 2006.

\bibitem{Sun2009}
Ke-Xun Sun, Saps Buchman, Robert Byer, Dan DeBra, John Goebel, Graham Allen,
  John~W Conklin, Domenico Gerardi, Sei Higuchi, Nick Leindecker, Patrick Lu,
  Aaron Swank, Edgar Torres, and Martin Trittler.
\newblock {Modular gravitational reference sensor development}.
\newblock {\em Journal of Physics: Conference Series}, 154:012026, 2009.

\bibitem{sun2011}
Ke-Xun. Sun, A.~Alfauwaz, M.~Alrufaydah, H.~Altwaijry, K.~Balakrishnan,
  S.~Buchman, R.~L. Byer, J.~W. Conklin, D.~B. DeBra, J.~Goebel, E.~Hultgren,
  and A.~Zoellner.
\newblock {Modular Gravitational Reference Sensor (MGRS) Technology
  Development}.
\newblock In {\em Proceedings of the 8th International LISA Symposium}, Journal
  of Physics Conference Series, 2011.

\bibitem{lange1964thesis}
B.~Lange.
\newblock {\em {The Control and use of Drag-free Satellites}}.
\newblock {PhD thesis}, Stanford University, 1964.

\bibitem{triad1974}
{Staff Of The Space Department}, {Staff Of The Guidance}, and {Control
  Laboratory}.
\newblock {A Satellite Freed of all but Gravitational Forces: ''TRIAD I''}.
\newblock {\em Journal of Spacecraft and Rockets}, 11:637, September 1974.

\bibitem{debra1999}
D.~B. {DeBra}.
\newblock {Design considerations for drag free satellites}.
\newblock In {W.~M.~Folkner}, editor, {\em Laser Interferometer Space Antenna,
  Second International LISA Symposium on the Detection and Observation of
  Gravitational Waves in Space}, volume 456 of {\em American Institute of
  Physics Conference Series}, pages 199--206, December 1998.

\bibitem{debra2011}
D.~B. {DeBra} and J.~W. {Conklin}.
\newblock {Measurement of drag and its cancellation}.
\newblock {\em Classical and Quantum Gravity}, 28(9):094015, May 2011.

\bibitem{higuchi2009}
S.~{Higuchi}, Ke-Xun. {Sun}, D.~B. {DeBra}, S.~{Buchman}, and R.~L. {Byer}.
\newblock {Design of a highly stable and uniform thermal test facility for MGRS
  development}.
\newblock {\em Journal of Physics Conference Series}, 154(1):012037, March
  2009.

\bibitem{alfauwaz2011}
A.~Alfauwaz and Ke-Xun. Sun.
\newblock {Design and Modeling of Highly Stable and Uniform Thermal Enclosure
  for Precision Space Experiment}.
\newblock In {\em Journal of Physics Conference Series}, Proceedings of the 8th
  International LISA Symposium, 2011.

\bibitem{Wass2005}
P~J Wass, H~M Ara{\'{u}}jo, D~N~a Shaul, and T~J Sumner.
\newblock {Test-mass charging simulations for the LISA Pathfinder mission}.
\newblock {\em Classical and Quantum Gravity}, 22(10):S311--S317, may 2005.

\bibitem{Araujo2005}
H~Araujo, P~Wass, D~Shaul, G~Rochester, and T~Sumner.
\newblock {Detailed calculation of test-mass charging in the LISA mission}.
\newblock {\em Astroparticle Physics}, 22(5-6):451--469, jan 2005.

\bibitem{Sumner2009}
T~J Sumner, DNA Shaul, M~O Schulte, S~Waschke, D~Hollington, and H~Ara\'{u}jo.
\newblock {LISA and LISA Pathfinder charging}.
\newblock {\em Classical and Quantum Gravity}, 26(9):094006, May 2009.

\bibitem{buchman1995}
S.~{Buchman}, T.~{Quinn}, G.~M. {Keiser}, D.~{Gill}, and T.~J. {Sumner}.
\newblock {Charge measurement and control for the Gravity Probe B gyroscopes}.
\newblock {\em Review of Scientific Instruments}, 66:120--129, January 1995.

\bibitem{Sun2006a}
Ke-Xun Sun, Brett Allard, Saps Buchman, Scott Williams, and Robert~L Byer.
\newblock {LED deep UV source for charge management of gravitational reference
  sensors}.
\newblock {\em Classical and Quantum Gravity}, 23(8):S141--S150, 2006.

\bibitem{Pollack2010}
S.~E. Pollack, M.~D. Turner, S.~Schlamminger, C.~A. Hagedorn, and J.~H.
  Gundlach.
\newblock {Charge management for gravitational-wave observatories using UV
  LEDs}.
\newblock {\em Physical Review D}, 81(2), jan 2010.

\bibitem{Olatunde2015}
Taiwo Olatunde, Ryan Shelley, Andrew Chilton, Paul Serra, Giacomo Ciani, Guido
  Mueller, and John Conklin.
\newblock {240 nm UV LEDs for LISA test mass charge control}.
\newblock {\em Journal of Physics: Conference Series}, 610:012034, may 2015.

\bibitem{Weber2003}
W.~J. Weber, L.~Carbone, A.~Cavalleri, R.~Dolesi, C.~D. Hoyle, M.~Hueller, and
  S.~Vitale.
\newblock {Possibilities for Measurement and Compensation of Stray DC Electric
  Fields Acting on Drag-Free Test Masses}.
\newblock sep 2003.

\bibitem{SET}
{Sensor Electronic Technology, http://www.s-et.com}.

\bibitem{Shaul2008}
D~N~a Shaul, H~M Araujo, G~K Rochester, M~Schulte, T~J Sumner, C~Trenkel, and
  P~Wass.
\newblock {Charge management for LISA and LISA Pathfinder}.
\newblock {\em International Journal of Modern Physics D}, 17(7):993--1003,
  2008.

\bibitem{Jiang1998}
Xinrong Jiang, C.~N. Berglund, Anthony~E. Bell, and William~A. Mackie.
\newblock {OS3: Photoemission from gold thin films for application in
  multiphotocathode arrays for electron beam lithography}.
\newblock {\em Journal of Vacuum Science {\&} Technology B: Microelectronics
  and Nanometer Structures}, 16(6):3374, Nov 1998.

\bibitem{Perl2006}
E~Perl.
\newblock {Test requirements for launch, upper stage, and space vehicles}.
\newblock Technical report, The Aerospace Corporation, September 2006.

\bibitem{Sun2009a}
Ke-Xun Sun, Nick Leindecker, Sei Higuchi, John Goebel, Sasha Buchman, and
  Robert~L Byer.
\newblock {UV LED operation lifetime and radiation hardness qualification for
  space flights}.
\newblock {\em Journal of Physics: Conference Series}, 154:012028, 2009.

\bibitem{Balakrishnan2011}
Karthik Balakrishnan, Eric Hultgren, John Goebel, and Ke-Xun Sun.
\newblock {Space Qualification Test Results of Deep UV-LEDs for AC Charge
  Management}.
\newblock In {\em 11th Spacecraft Charging Technology Conference}, 2011.

\bibitem{Newport2010}
{Model 918D Series User's Manual}.
\newblock Technical report, Newport Corporation, 2010.

\bibitem{mcmaster}
McMaster-Carr.
\newblock personal communication, 2012.

\bibitem{Engineering2011}
Assistant Secretary of Defense for~Research and Engineering.
\newblock {Technology Readiness Assessment (TRA) Guidance}.
\newblock Technical Report May, Department of Defense, 2011.

\bibitem{kosmotras2014}
{Kosmotras, 19 June 2014. Dnepr Cluster Mission 2014.
  http://www.kosmotras.ru/en/launch15/}.

\bibitem{Instruments2014}
Texas Instruments.
\newblock {LMP7721 3-Femtoampere Input Bias Current Precision Amplifier
  Datasheet}.
\newblock Technical report, Dallas, Texas, 2014.

\bibitem{Jackson1999}
John~David Jackson.
\newblock {\em {Classical electrodynamics}}.
\newblock Wiley, 1999.

\end{thebibliography}

\end{document}